# From rankings to funnel plots: the question of accounting for uncertainty when assessing university research performance[1]

Giovanni Abramo (*corresponding author*)
  *Laboratory for Studies of Research and Technology Transfer*
  *Institute for System Analysis and Computer Science (IASI-CNR)*
  *National Research Council of Italy*
    ADDRESS: Istituto di Analisi dei Sistemi e Informatica, Consiglio Nazionale delle Ricerche
    Via dei Taurini 19, 00185 Roma - ITALY
    tel. +39 06 7716417, fax +39 06 7716461, giovanni.abramo@uniroma2.it

Ciriaco Andrea D'Angelo
  *University of Rome 'Tor Vergata' and Institute for System Analysis and Computer Science-National Research Council of Italy*
    ADDRESS: Dipartimento di Ingegneria dell'Impresa, Università degli Studi di Roma 'Tor Vergata'
    Via del Politecnico 1, 00133 Roma - ITALY
    Tel. and fax +39 06 72597362, dangelo@dii.uniroma2.it

Leonardo Grilli
  *University of Florence – Italy*
    ADDRESS: Dipartimento di Statistica, Informatica, Applicazioni 'G. Parenti', Università degli Studi di Firenze
    Viale Morgagni, 59, 50134 Firenze - ITALY
    Tel. +39 055 2751552 - fax +39 055 4223560, grilli@disia.unifi.it

**Abstract**
The work applies the funnel plot methodology to measure and visualize uncertainty in the research performance of Italian universities in the science disciplines. The performance assessment is carried out at both discipline and overall university level. The findings reveal that for most universities the citation-based indicator used gives insufficient statistical evidence to infer that their research productivity is inferior or superior to the average. This general observation is one that we could indeed expect in a higher education system that is essentially non-competitive. The question is whether the introduction of uncertainty in performance reporting, while technically sound, could weaken institutional motivation to work towards continuous improvement.

---

# 1. Introduction

In the current knowledge-based economy, research and higher education systems play a significant role in supporting the competitiveness and socio-economic growth of nations, through the education of white collar workers and production of new knowledge. Improvement in the research and higher education infrastructure has with good reason become a policy priority for a growing number of governments. Among the interventions to increase the effectiveness and efficiency of research institutions, and the socio-economic returns from public spending on R&D, a growing number of countries are launching national assessment exercises of their research institutions. The exercises are intended to accomplish various aims, as selected by the national governments: for informing selective funding of research institutions; stimulating better research performance; reducing information asymmetry between supply and demand in the market for knowledge; informing policy formulation and strategic decisions; and last but not least, demonstrating that public investment in research is effective and delivers public benefits. An international comparative analysis of performance-based research funding (Hicks, 2012), indicates that subsequent to the example of the original UK research assessment exercise (the RAE, in 1986), at least 14 other countries (China, Australia, New Zealand and 11 in the EU) have chosen to implement national assessment exercises as the basis for directing at least some portion of public financing for research institutions. Alongside this, several annual world university rankings continuously receive great media attention, influencing opinion and practical choices. The various national and international assessments employ a variety of indicators and methods (bibliometric, peer-review, informed peer-review, surveys) to assess institutions' research performance. A common feature of the vast majority of the performance assessments is the lack of confidence intervals indicating the likely range of population values. The scores observed for research performance are in fact related to the assumptions and limits of the particular measurement method and indicators, and in the case of aggregate measures to the different sizes of the research institutions. Accounting for the uncertainty embedded in the measurements is crucial to establish whether the performance of an institution is truly outstanding or the result of random fluctuations.

Assessment of research performance is affected by both bias and uncertainty factors[2]. Bias factors generate fluctuations with systematic effects. A typical source of bias is the differing intensity of publication and citation across fields, which the evaluator ideally attempts to limit through a fine-grained classification of scientists and field-normalization of citations. Conversely, uncertainty factors randomly affect the assessment, meaning they will generate fluctuations without systematic effects in favor or against particular groups. Typical factors increasing uncertainty in performance assessment are the variability in intensity of production due to personal events, or due to the patterns characteristic of research projects, or the varying lengths of review and publication time across journals. Ideally, uncertainty factors should again be limited, but they cannot be completely eliminated. Notwithstanding uncertainty, the analyses of research performance can still be valid, as long as the reporting includes measures of uncertainty. The recent Leiden manifesto (Hicks et al., 2015) wisely recommends that practitioners 'avoid misplaced concreteness and false precision' in reporting performance values, and that 'if uncertainty and error can be quantified, for instance using error bars, this information should accompany published indicator values'.

---

[2] We refer the reader to our previous work (Abramo et al., 2015) for a detailed analysis of all factors of uncertainty and bias in research performance assessment.

However, indications of uncertainty are generally not provided for the popular international 'league tables' of universities. This is true whether the performance scores and relevant rankings are produced by 'non-bibliometricians', such as the Shanghai Jiao Tong University Ranking (SJTU, 2016), QS World University Rankings (QS, 2016) and Times Higher Education World University Rankings (THE, 2016), or whether they are produced by bibliometricians themselves, such as the Scimago Institutions Ranking (Scimago, 2016). The CWTS Leiden Rankings few years ago integrated stability intervals (Waltman et al., 2012). In our studies concerning Italian university research productivity rankings (e.g. Abramo et al., 2011), we have ourselves not usually provided the likely range of performance values. Recently we attempted to deal with this shortcoming, by applying a funnel plot methodology to measure and visualize uncertainty in the research performance of the institutions. The funnel plot shows the uncertainty in data values by adding confidence bands, indicating the range where research performance indicator's values are expected to lie on the basis of the institution's size. To illustrate the funnel plot methodology, we applied it to measure uncertainty in the research productivity of Italian universities active in Biochemistry (Abramo et al., 2015). The results showed that just one university out of 42 had truly outstanding research performance, while for 79% of universities the performance was not different from the overall mean, at a 5% significance level. Should the results in Biochemistry be confirmed for all sciences, then any performance rankings neglecting uncertainty would be misleading for policy and decision-making.

Considering this question and its implications, the current work thus extends the application of funnel plots to measure uncertainty in all fields and disciplines of the sciences. The goal is to identify the proportion of outstanding universities in each single field of research (192 fields), namely units whose difference from the overall mean is statistically significant. The analyses will be carried out separately for each discipline (nine disciplines).

The provision of reliable institutional research performance scores, including visualization of uncertainty levels, has implications for both stakeholders and policy makers. The stakeholders can include anyone who draws on or is influenced by the rankings, from the casual observer to the potential student, to the interested enterprise and the highest political levels. Suffice it to think of the many countries that allocate public funding according to the rankings stemming from national research assessment exercises. Or how in the Italian case, parliament recently considered a proposal to normalize the graduation scores of candidates competing for public positions, by the 'quality' score of their degree-granting university.

We refer the reader to our previous work (Abramo et al., 2015) for an overview of the quite limited literature on measuring uncertainty in research performance, as well as a description of the funnel plot methodology. We would like to add here a work by Claassen (2015), which was published meanwhile. The author measures uncertainty in university quality estimates by eight different world ranking systems, showing that the difference between universities ranked $50^{th}$ and $100^{th}$, and $100^{th}$ and $250^{th}$, is not significant.

The funnel plot methodology presents advantages over other methods for visualizing uncertainty. For example, in the popular caterpillar plot the performance assessment of the units are plotted in increasing order and endowed with confidence intervals (see Spiegelhalter, 2005, and the references therein). The lengths of the intervals summarize the uncertainty and a unit whose interval is above (below) zero is judged to have a performance significantly above (below) the overall mean. Even if a caterpillar plot is technically correct, it may not be effective in communicating the results because: i) it does not explicitly show the relationship between the level of uncertainty and the volume or size of the units, and ii) it leads the reader towards undue emphasis on the ranking of the units though the reliability of the ranking is not assessed (the exact position of a unit is often found to be highly uncertain). The funnel plot overcomes these limitations. In fact, it shows the uncertainty in data values by adding confidence bands, indicating the range where research performance indicator's values are expected to lie on the basis of the institution's size. The visualization of uncertainty is useful in both analyzing the data and communicating the results. Nevertheless, applying the funnel plot to research performance assessment may entail visualization problems when it comes to display a large number of research institutions, as it may occur in some countries.

In the next section we present the dataset and the research performance indicator used in the current analysis. In Section 3 we show the results from applying the funnel plot to measure uncertainty in the research performance of the Italian universities in nine science disciplines. Section 4 offers the conclusions.

## 2. Methods and data

### 2.1 The funnel plot graphical display

Bibliometric assessment of performance is based on countable research outputs (impacts), and in some cases also on inputs. In addition to the uncertainty inherent in the measurement of outputs and inputs, there is also the further issue that bibliometricians compare institutions on the basis of the average of the measured performance of their researchers. In this context, accounting for uncertainty involves the additional feature that the various uncertainty factors are aggregated at the institution level, so that the amount of uncertainty will be inversely related to the size. Indeed in all rankings, small organizations can often fall at the extremes as a consequence of high variability, while large ones are instead generally situated around the middle, as a consequence of their low aggregated variability.[3] This differential variability due to size can be effectively handled by the funnel plot methodology.

The funnel plot is a graphic display for visualizing the uncertainty in the performance assessment of units as a function of their volume or size. It was originally developed in meta-analysis and later adapted to the comparison of institutions with different volumes, such as hospitals (Spiegelhalter, 2005; Ieva and Paganoni, 2015). The funnel plot has two elements: i) a scatter of institutional outcome (in our case, the institution's research performance) against size (number of researchers); ii) confidence bands around the overall mean to assess if the observed outcome is statistically significant at a given level (e.g. 5%). As the size of the institution increases, the standard error of the outcome decreases, thus the confidence bands converge toward the overall mean of the outcome. Typically, the institution's outcome is the mean of the chosen performance indicator at the individual level, thus the standard error and the implied confidence bands are inversely proportional to the square root of the number of observations (size), yielding funnel-shaped bands. The funnel plots usually show most institutions as falling within the bands, meaning there is no evidence that their performance is anomalous (and also implying that rankings would be misleading). Attention should instead be focused on those institutions falling outside the bands, whose performance is likely to be truly outstanding and worthy of closer scrutiny.

---

[3]It should be noted that this effect is not a consequence of varying returns to size or scope: in fact, it has been shown that in general there are no notably varying returns to scale (Abramo et al., 2012a) and to scope of research (Abramo et al., 2014).

We refer the reader to our previous work (Abramo et al., 2015) for the calculations underlying the construction of the funnel plots. We use the statistical software Stata 13, however the steps for constructing a funnel plot are so simple that they could also be implemented using a spreadsheet.

**2.2 The research performance indicator**

We depart from the mainstream and contend that all size-independent indicators based on the ratio to publications, such as the world-renowned Mean Normalized Citation Score, or MNCS, (Waltman et al., 2011) are invalid indicators of performance (Abramo & D'Angelo, 2016a, and 2016b). We measure the research performance by an indicator of productivity. Most bibliometricians define productivity as the number of publications of the unit in the period under observation. Because publications have different values (impact), we adopt a more meaningful definition of productivity: the value of output per unit value of labor, all other production factors being equal. The latter recognizes that the publications embedding new knowledge have a different value or impact on scientific advancement, which can be approximated with citations. Because citation behavior varies by field, we standardize the citations for each publication with respect to the average of the distribution of citations for all the cited Italian publications indexed in the same year and the same WoS subject category.[4] Furthermore, research projects frequently involve a team, which is registered in the co-authorship of publications. In this case we account for the fractional contributions of scientists to outputs which is, in the case of the life sciences, further signaled by the position of the authors in the list of authors. When measuring labor productivity, if there are differences in the production factors (scientific instruments, materials, databases, support staff, etc.) available to each scientist, then normalization should be conducted. Unfortunately, relevant data at the individual level are not generally available. Thus a necessary assumption is often that the resources available to single scientists within the same field are the same. A further assumption, again unless specific data are available, is that the hours devoted to research are more or less the same for each individual. However, one available source of information about input is the average salary per academic rank. In the Italian university system, all professors of the same academic rank and seniority receive the same salary, regardless of the university that employs them. The information on individual salaries is unavailable but the salaries ranges for rank and seniority are published. Thus we have approximated the salary for each individual as the national average of their academic rank. Failure to account for the cost of labor would result in ranking distortions, because it favors universities with a higher share of full professors, as shown by Abramo et al. (2010).

---
[4]Abramo et al. (2012a) demonstrated that the average of the distribution of citations received for all cited publications of the same year and subject category is the best-performing scaling factor.

At the individual level, we measure the average yearly productivity, termed the fractional scientific strength (*FSS*), as follows:[5]

$$FSS = \frac{1}{w_R} \cdot \frac{1}{t} \sum_{i=1}^{N} \frac{c_i}{\bar{c}} f_i \qquad [1]$$

where the symbols are defined as follows:
$w_R$ = average yearly salary of the professor[6]
$t$ = number of years of work by professor in period under observation
$N$ = number of publications by professor in period under observation
$c_i$ = citations received by publication *i*
$\bar{c}$ = average of distribution of citations received for all cited publications indexed in same year and subject category of publication *i*
$f_i$ = fractional contribution of professor to publication *i*.

Fractional contribution equals the inverse of the number of authors, in those fields where the practice is to place the authors in simple alphabetical order, but assumes different weights in other cases. For the life sciences, widespread practice in Italy and abroad is for for the authors to indicate the various contributions to the published research by the order of the names in the byline. If the first and last authors belong to the same university, 40% of the citation is attributed to each of them, the remaining 20% is divided among all other authors. If the first two and last two authors belong to different universities, 30% of the citation is attributed to the first and last authors, 15% of the citation is attributed to the second and penultimate, the remaining 10% is divided among all others[7]. Failure to account for the number and position of authors in the byline would result in notable ranking distortions both at the individual (Abramo et al. 2013a), and aggregate (Abramo et al. 2013b) levels.

**2.3 Data**

In the Italian university system all professors are classified in one and only one field, named Scientific Disciplinary Sector (SDS), 370 in all. 193 SDSs fall in Science, while the remainder in Arts and Humanities, and Social Sciences. The SDSs are grouped into 14 disciplines, named University Disciplinary Areas. 9 UDAs fall in Science. We assess the research performance of universities in Science, at the SDS, UDA and overall university levels. We restrict our analysis to Science because we use a bibliometric indicator of research performance, and in Science the publications indexed in bibliometric databases are considered a satisfactory proxy of overall research output (Moed, 2005).

---

[5] A more extensive theoretical dissertation on how to operationalize the measurement of productivity can be found in Abramo and D'Angelo (2014).
[6] We adopt the following salary normalization coefficients: 1 for assistant; 1.4 for associate; 2 for full professors (source DALIA - https://dalia.cineca.it/php4/inizio_access_cnvsu.php, last accessed on July 5, 2016).
[7] The weightings were assigned on the basis of advice from senior Italian professors in the life sciences. The values could be changed to suit different practices in other national contexts.

We first measure the productivity of each professor, and then average the productivity values of the faculty at each university at the SDS, UDA and overall university levels. The period of research production analyzed is 2008-2012. Citations are counted on May 15, 2014. The citation window is large enough to assure an adequate estimate of the impact of each publication (Abramo et al., 2011). For reliable assessement of research performance, any index should be calculated over a sufficiently long period (Abramo et al., 2012b), thus we excluded those professors with less than three years on faculty over the observed period.

We have extracted data on the faculty at each university from the database on Italian university personnel, maintained by the Italian Ministry of Education, University and Research[8]. Such database provides information on the affiliation, SDS, and academic rank of each professor in Italy. Our dataset is made of 35,926 professors in the SDSs falling in the 9 Science UDAs, on staff in 64 universities.

The scientific production was extracted from the Italian Observatory of Public Research, a bibliometric database developed and maintained by the first two authors and derived under license from the Thomson Reuters WoS. Beginning from the raw data of the WoS, and applying a complex algorithm to reconcile the author's affiliation and disambiguation of the true identity of the authors, each publication (article, review and conference proceeding) is attributed (3% error - harmonic average of precision and recall) to the university professor or professors that produced it (D'Angelo et al., 2011).

## 3 University research performance

In this section we analyze university productivity using the funnel plot methodology. We start from the field level (SDS) within a discipline (UDA). Because of space limitations, we show the plots for the SDSs of only one UDA, namely Physics. Following this example, we then present the plots for each of the nine science UDAs.

To measure uncertainty, similarly to Abramo et al. (2015), we use funnel plots with Normal-based bands, namely $\bar{y} \pm z_{\alpha/2} s/\sqrt{n_j}$, where $\bar{y}$ is an estimate of the overall mean of the FSS index (possibly transformed to improve normality), $s$ is an estimate of the standard deviation, $n_j$ is the number of professors of institution $j$, and $z_{\alpha/2}$ denotes the value of the Normal distribution with probability $\alpha/2$ on the right tail[9]. Normal-based bands are valid if the university means $\bar{y}_j$ are approximately normally distributed, which is likely for large universities due to the Central Limit Theorem. In practice, for a satisfactory approximation to normality we exclude universities where the number of professors in the observed SDS or UDA falls below a given threshold: 5 for analysis at the SDS level, 10 at the UDA level. Furthermore, we apply the zero-skewness log transform ln(FSS+$k$), where the value of $k$ is selected so that the distribution of the transformed data is symmetric (Box and Cox, 1964). We evaluate the normality of the distribution of university means with both a statistical test and a graphical display, namely the normal quantile plot. Due to space limitations, in this paper we only report the *p*-value of the statistical test, specifically the Shapiro-Wilk test (Royston, 1992) applied before and after the transform.

---

[8]http://cercauniversita.cineca.it/php5/docenti/cerca.php, last accessed on July 5, 2016.
[9] We draw two pairs of bands: internal bands with $z_{\alpha/2}$=2 corresponding to a confidence level of about 95%, and external bands with $z_{\alpha/2}$=3 corresponding to a confidence level of about 99.7%.

**3.1 Field level analysis**

Table 1 summarizes the data for the analysis at the SDS level, with reference to the seven fields of Physics distinguished under the Italian system. As discussed above, for each field the data refer to universities with at least five professors. Comparing the median and maximum number of professors in each SDS, it is clear that the universities are widely different in terms of size. The estimated parameter $k$ of the zero-skewness log transform varies across fields, although the differences are modest. Using a 5% level, the Shapiro-Wilk test rejects normality of the university means on the original data for three SDSs, whereas it does not reject normality of the means on the transformed data with the only exception of FIS/03. Figure 1 shows the funnel plots for the SDSs of Physics, excluding FIS/06 which is too small for a meaningful analysis. In all fields except FIS/05, there are few universities outside the bands, meaning in positions indicating noteworthy performances. For example, FIS/02 shows one university below the 3SD band (very poor performance), whereas FIS/03 shows two universities above the 3SD band (excellent performance). However, most universities lie within the bands. The implied rankings should thus be interpreted with great caution, since the positions are quite uncertain, and in most instances there is not enough evidence for claims of superiority or inferiority.

*Table 1: Data for the analysis at SDS level for the UDA Physics*

| SDS* | N. of Universities§ | Professors (total) | Professors (median) | Professors (max) | Shapiro-Wilk *p*-value (original data) | $k$ (transform constant) | Shapiro-Wilk *p*-value (transformed data) |
|---|---|---|---|---|---|---|---|
| FIS/01 | 40 | 901 | 18 | 70 | 0.006 | -0.039 | 0.234 |
| FIS/02 | 26 | 319 | 11.5 | 33 | 0.576 | -0.042 | 0.257 |
| FIS/03 | 33 | 439 | 12 | 31 | 0.000 | -0.051 | 0.006 |
| FIS/04 | 16 | 113 | 6 | 12 | 0.069 | -0.032 | 0.586 |
| FIS/05 | 16 | 164 | 8 | 28 | 0.934 | -0.042 | 0.497 |
| FIS/06 | 4 | 32 | 8 | 9 | 0.487 | -0.011 | 0.577 |
| FIS/07 | 28 | 288 | 9 | 20 | 0.040 | -0.026 | 0.901 |

* FIS/01=Experimental Physics; FIS/02=Theoretical Physics, Mathematical Models and Methods; FIS/03=Material Physics; FIS/04=Nuclear and Subnuclear Physics; FIS/05=Astronomy and Astrophysics; FIS/06=Physics for Earth and Atmospheric Sciences; FIS/07=Applied Physics (Cultural Heritage, Environment, Biology and Medicine)

§ with at least 5 professors in the SDS

*Figure 1: Funnel plots of research productivity (average transformed FSS) of Italian universities with at least 5 professors in each SDS of Physics, over the 2008-2012 period*

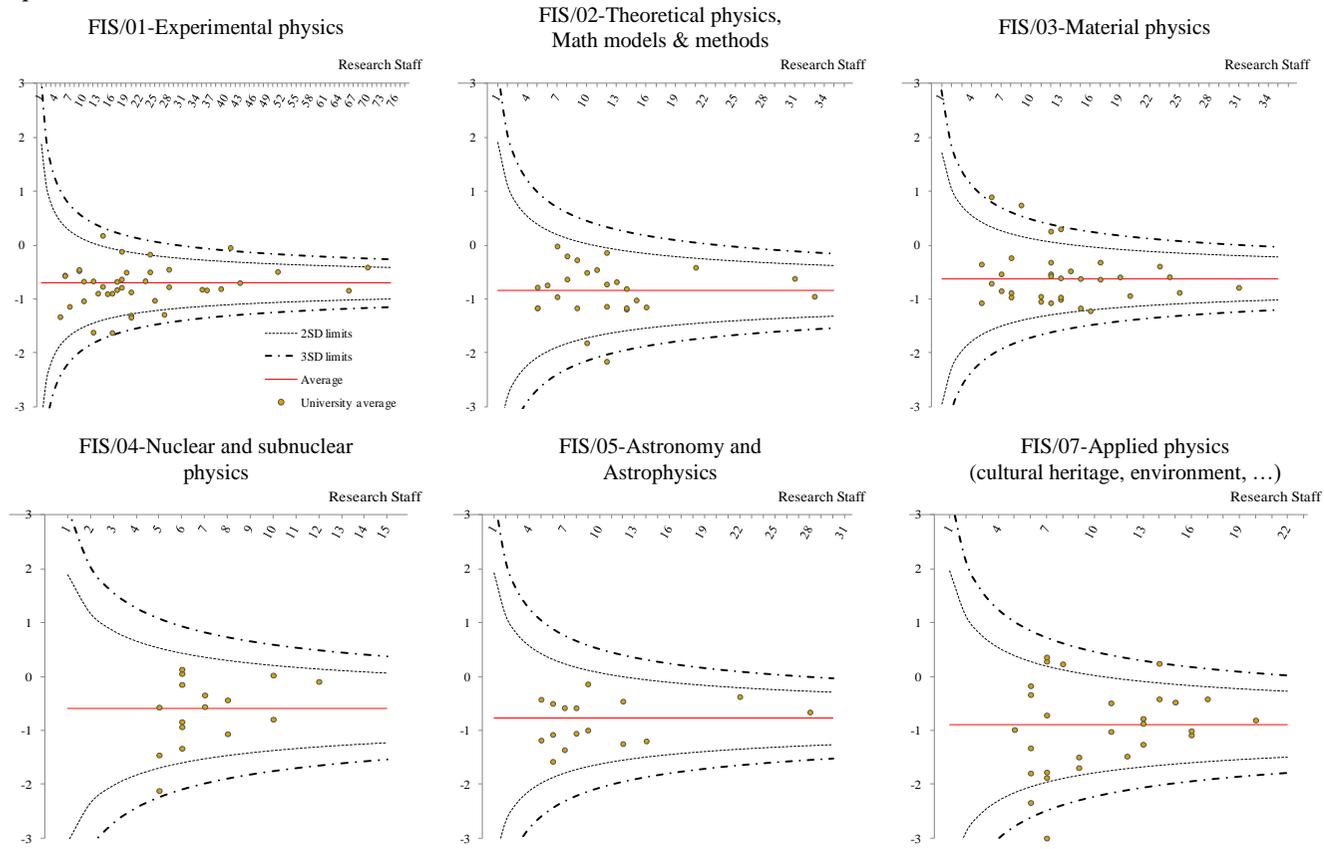

## 3.2 Discipline level analysis

In this subsection, we present the analysis at the UDA level. We exclude universities with less than 10 professors in the observed UDA. Table 2 summarizes the data for the analysis for each UDA. According to the Shapiro-Wilk test, the zero-skewness transform improves the normality of the university means[10] except in the case UDA 1 (Mathematics and computer science). A look at the distributions of the university means in the original scale suggests that the normality test fails due to asymmetry, namely there is a long right tail generated by a few excellent universities.[11] For each UDA, Table 3 provides the pairs of summary statistics needed for the construction of the funnel plot, namely the overall mean ($\bar{y}$) and the within-university standard deviation ($s$). The values cannot be compared across UDAs, since they are computed on different scales. Indeed, the constant $k$ of the zero-skewness transform is different for each UDA. However, this is not a cause of concern, since the underlying bibliometric methodology is devised to perform comparisons only within UDAs, and we consistently avoid any comparisons across UDAs. Figure 2 shows the funnel plots for the nine UDAs under consideration. For more compact diagramming, the horizontal scale in the different graphs is related to the maximum size observed in each UDA.

*Table 2: Data for the analysis at UDA level*

| UDA[†] | N. of Universities[§] | Professors (total) | Professors (median) | Professors (max) | Shapiro-Wilk $p$-value (original data) | $k$ (transform constant) | Shapiro-Wilk $p$-value (transformed data) |
|---|---|---|---|---|---|---|---|
| MAT | 50 | 3,331 | 49.5 | 232 | 0.171 | -0.037 | 0.035 |
| PHYS | 43 | 2,422 | 50 | 165 | 0.004 | -0.095 | 0.405 |
| CHEM | 44 | 3,107 | 60 | 242 | 0.000 | -0.089 | 0.242 |
| EAS | 32 | 1,129 | 32 | 76 | 0.000 | -0.089 | 0.315 |
| BIO | 52 | 5,150 | 75 | 345 | 0.025 | -0.060 | 0.939 |
| MED | 43 | 10,903 | 200 | 1,333 | 0.243 | -0.026 | 0.444 |
| AGR | 29 | 3,125 | 80 | 284 | 0.384 | -0.059 | 0.623 |
| CENG | 36 | 1,595 | 34 | 151 | 0.014 | -0.019 | 0.393 |
| IENG | 47 | 5,164 | 76.5 | 607 | 0.043 | -0.072 | 0.311 |

[†]*MAT=Mathematics and computer science; PHYS=Physics; CHEM=Chemistry; EAS=Earth sciences; BIO=Biology; MED=Medicine; AGR=Agricultural and veterinary sciences; CENG=Civil engineering; IENG=Industrial and information engineering*
[§] *with at least 10 professors in the UDA*

---

[10] Note that the zero-skewness log-transform is applied to the data at the individual level, thus it does not necessarily eliminate the skewness of the university means.
[11] UDA 1 is an exception. Nonetheless to aid in the consistent interpretation of results, we also transform the data of UDA 1.

*Table 3: Summary statistics for the funnel plots at UDA level (transformed data)*

| UDA[†] | Overall mean | Within-university standard deviation |
|---|---|---|
| MAT | -1.2227 | 1.5173 |
| PHYS | -0.5485 | 1.0898 |
| CHEM | -0.4700 | 1.0124 |
| EAS | -0.6280 | 1.1249 |
| BIO | -0.7249 | 1.2249 |
| MED | -1.2827 | 1.6385 |
| AGR | -0.9166 | 1.3311 |
| CENG | -1.6596 | 1.8240 |
| IENG | -0.7877 | 1.2357 |

[†]*MAT=Mathematics and computer science; PHYS=Physics; CHEM=Chemistry; EAS=Earth sciences; BIO=Biology; MED=Medicine; AGR=Agricultural and veterinary sciences; CENG=Civil engineering; IENG=Industrial and information engineering*

*Figure 2: Funnel plots of research productivity (average transformed FSS) of Italian universities with at least 10 professors in each UDA, over the 2008-2012 period*

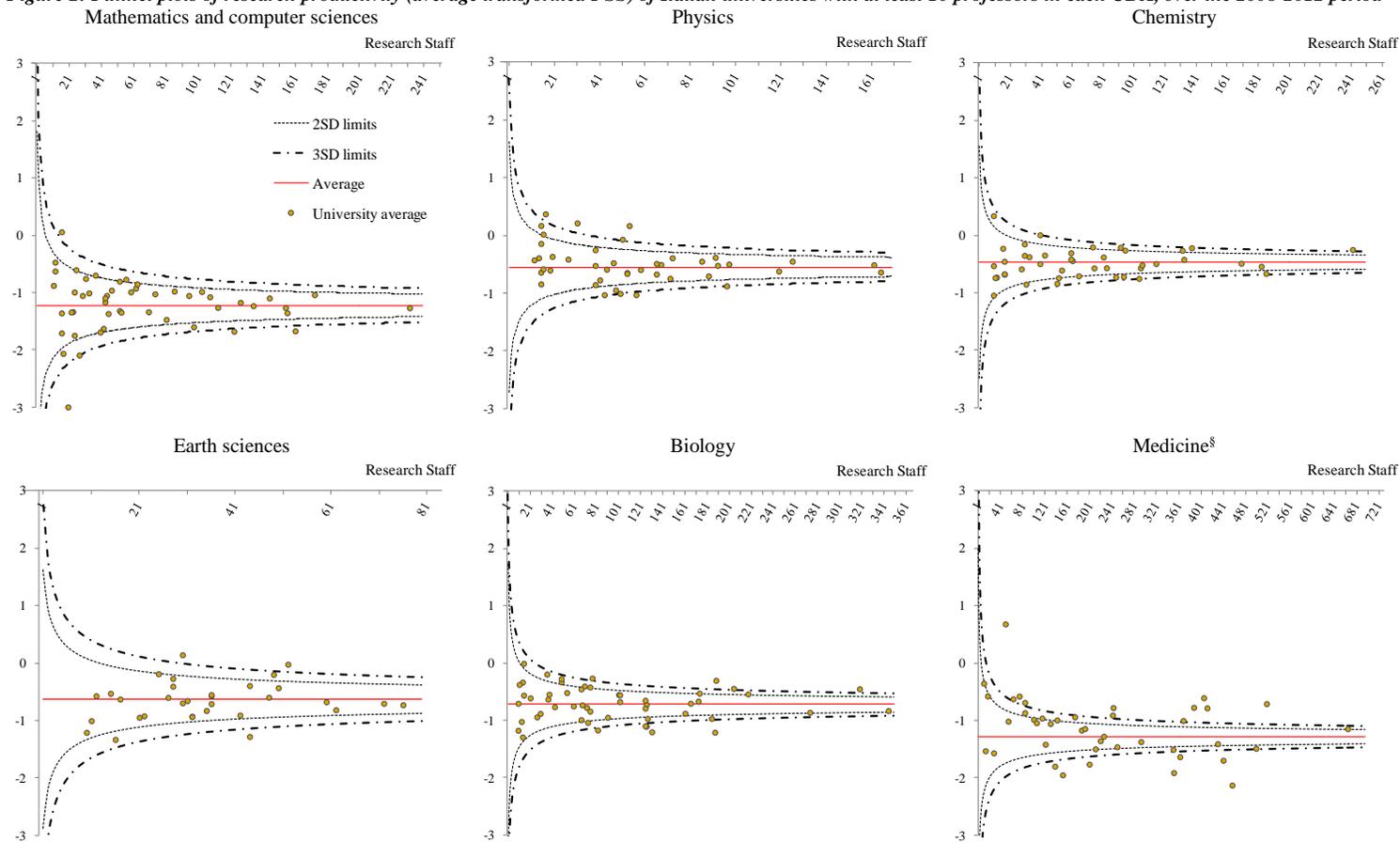

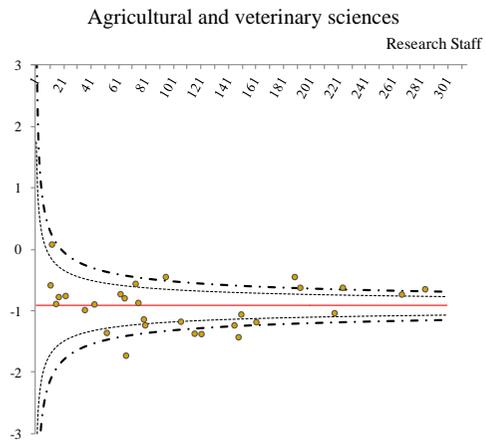 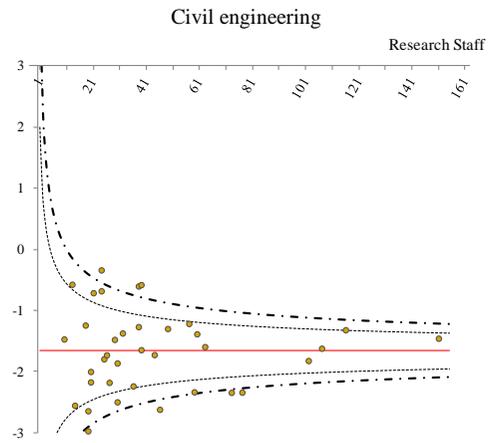 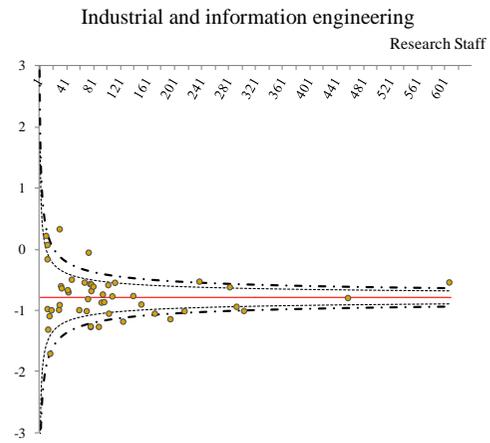

*§ University of Rome 'Sapienza' has been omitted because the size of its research staff is it out of scale.*

In all the UDAs there are few universities outside the bands. Medicine has the largest number and percentage of outliers (19 out of 43, i.e. 44%). This fact could be ascribed to the specific features of the discipline: for example, research in medicine is highly dependent on resources and institutional contexts,[12] which are quite different across the country. However, it is more likely that the large number of outliers in Medicine has a statistical explanation. Medicine is by far the largest UDA with a median of 200 professors per university (Table 2), which favors the detection of outliers. Specifically, as discussed in Abramo et al. (2015), the funnel plot amounts to a series of tests of the null hypothesis that the mean performance of the institution under consideration is equal to the overall mean performance. The Type I error rate is fixed at a level determined by the bands (e.g. about 5% using two-SD bands), but the Type II error rate is inversely related to the size of the institution, namely the number of professors. In other words, the probability of correctly detecting an institution as an outlier is greater for large-sized institutions. To check this behavior in our application, Table 4 reports the number of outlying universities for each UDA, both overall and by size intervals. The last row of the table reveals that the percentage of outlying institutions indeed increases with the size of the institutions. Note that 14 out of 19 assessed units sized over 300 are in Medicine, thus it is not surprising that this UDA has the largest number of outlying universities.

*Table 4: Stratification of outliers by range of institutional size (in brackets the numbers of universities below the lower and above the higher three-SD bands)*

| UDA[†] | N. of Universities[§] | Research staff | | | |
|---|---|---|---|---|---|
| | | [10;100] | (100;200] | (200;300] | Over 300 |
| MAT | 50 (3;1) | 38 (1;1) | 11 (2;0) | 1 (0;0) | |
| PHYS | 43 (1;4) | 39 (1;4) | 4 (0;0) | | |
| CHEM | 44 (1;1) | 33 (1;0) | 10 (0;0) | 1 (0;1) | |
| EAS | 32 (1;2) | 32 (1;2) | | | |
| BIO | 52 (5;4) | 30 (1;1) | 17 (4;1) | 3 (0;1) | 2 (0;1) |
| MED | 43 (8;11) | 11 (0;4) | 11 (3;0) | 7 (0;2) | 14 (5;5) |
| AGR | 29 (4;5) | 16 (1;1) | 9 (3;2) | 4 (0;2) | |
| CENG | 36 (4;3) | 32 (4;3) | 4 (0;0) | | |
| IENG | 47 (6;4) | 31 (3;2) | 9 (2;0) | 4 (0;1) | 3 (1;1) |
| Total obs. | 376 (33;35) | 262 (13;18) | 75 (14;3) | 20 (0;7) | 19 (6;7) |
| % outlying | 18,1% | 11,5% | 22,7% | 35,0% | 68,4% |

[†] *MAT=Mathematics and computer science; PHYS=Physics; CHEM=Chemistry; EAS=Earth sciences; BIO=Biology; MED=Medicine; AGR=Agricultural and veterinary sciences; CENG=Civil engineering; IENG=Industrial and information engineering*
[§] *with at least 10 professors in the UDA*

Finally, in order to assess the relationships between disciplines, Table 5 reports the Spearman correlations of the productivity measured by FSS between the nine UDAs. Except for three instances, all the correlations are positive. They are also generally moderate, with the noteworthy and meaningful exception the 0.713 correlation between UDAs 5 (Biology) and 6 (Medicine). A consequence of positive correlations is that the aggregation of the disciplines to arrive at institutional scores amplifies the differences between the universities. This fact then raises the probability of detecting outliers at the overall university level, as it reinforces the greater statistical power gained by larger institutional size. However, stakeholders in higher education should in theory be more interested in comparing university performance per discipline or field of research.

---
[12] For instance, performance in clinical research may benefit from large numbers of patients.

Indeed, it is at those levels that stakeholders could normally make meaningful or effective decisions. We therefore purposely omit the presentation of a funnel plot at the overall university level.

*Table 5: Spearman correlations of the productivity measured by FSS between the nine UDAs of Italian universities, over the 2008-2012 period.*

| UDA[†] | 1 | 2 | 3 | 4 | 5 | 6 | 7 | 8 | 9 |
|---|---|---|---|---|---|---|---|---|---|
| MAT | - | 0.291 | 0.357 | -0.147 | 0.133 | 0.191 | -0.410 | 0.259 | 0.353 |
| PHYS | 0.291 | - | 0.076 | 0.134 | 0.358 | 0.436 | 0.360 | -0.128 | 0.411 |
| CHEM | 0.357 | 0.076 | - | 0.208 | 0.146 | 0.477 | 0.196 | 0.370 | 0.368 |
| EAS | -0.147 | 0.134 | 0.208 | - | 0.543 | 0.533 | 0.170 | 0.133 | 0.028 |
| BIO | 0.133 | 0.358 | 0.146 | 0.543 | - | 0.713 | 0.504 | 0.314 | 0.349 |
| MED | 0.191 | 0.436 | 0.477 | 0.533 | 0.713 | - | 0.545 | 0.083 | 0.383 |
| AGR | -0.410 | 0.360 | 0.196 | 0.170 | 0.504 | 0.545 | - | 0.486 | 0.586 |
| CENG | 0.259 | -0.128 | 0.370 | 0.133 | 0.314 | 0.083 | 0.486 | - | 0.500 |
| IENG | 0.353 | 0.411 | 0.368 | 0.028 | 0.349 | 0.383 | 0.586 | 0.500 | - |

[†] MAT=Mathematics and computer science; PHYS=Physics; CHEM=Chemistry; EAS=Earth sciences; BIO=Biology; MED=Medicine; AGR=Agricultural and veterinary sciences; CENG=Civil engineering; IENG=Industrial and information engineering

**5. Conclusions**

The funnel plot methodology is unlike traditional research performance rankings in that it allows the measurement and visualization of uncertainty. The 376 observations of Italian universities in the nine science disciplines show that in only 18% of cases there is evidence that institutional research performance is above or below the mean. Leaving aside the discipline of Medicine, this occurs in only 13% of cases. In the discipline of Chemistry, for 95% of universities there is insufficient evidence for any claims of superiority or inferiority. Overall, most Italian universities show a research performance that is not statistically different from the overall mean. This scarce differentiation between universities could well be expected, as the Italian higher education system is essentially non-competitive (Abramo et al. 2012c). It would be interesting to compare the results from Italy with those from more competitive higher education systems, such as the US or the UK.

The implications of applying the funnel plot methodology to the assessment of research performance seem relevant to the policy and decision-making levels. The production of rankings mistakenly transmits the idea of a definitive position for each institution within the population, whereas funnel plots indicate which institutions are very likely to have an underlying performance above or below the overall mean. For the large share composed of the remaining institutions, the relative positions with respect to the mean are highly uncertain. Compared to the view afforded by simple rankings, the funnel plot representation of research performance could induce different choices by stakeholders, for instance in the procedures for selective allocation of public funding. Based on the past two Italian research assessment exercises (2001-2003 VTR, 2006-2010 VQR), a share of the annual funding allocation for universities was distributed according to performance rankings, which did not account for uncertainty. The application of the current 2011-2014 VQR will follow suit. To the best of our knowledge, in other countries as well, performance-based funding allocations consistently refer to scores that do not account for uncertainty.

Taking a technical perspective, bibliometricians have stated that performance reporting should always include measures of uncertainty. However, from a policy and managerial view, consideration should be given to the potential counter effects of such measures on stakeholders. Indeed, the adoption of research assessment exercises and performance reporting seems to have stimulated a culture of continuous improvement, which is probably the main aim envisaged in such assessments (Marginson, 1997). The question is whether the introduction of uncertainty in performance reporting could weaken the various effects at the basis of continuous improvement. Given the publication of measures of uncertainty, would the then undistinguished Chemistry faculties of 95% of Italian universities willingly respond to the incentives for improvement in ratings, to the extent that they do under 'definitive' ranking? The risk is that performance reporting visualizing uncertainty might weaken the motivational power embedded in rankings. What effect would such measures have on the case of prospective top students, who resort to the rankings to select their places of study, or on faculty seeking to work at universities moving towards better performance? The uncertainty visualized in the funnel plots might easily transfer to the many stakeholders. At that point, the effects on their various choices might be unpredictable. For instance, a question arises on how policy makers can use confidence intervals to allocate funding. If the approach is to allocate funding on a continuous scale, then the actual value of the performance score is still the best estimate, regardless of the level of uncertainty. However, for small universities this approach may produce large variations of funding from year to year, so one could envision a more 'conservative' approach, where the research performance score is used to reward or penalize especially outlying universities (i.e. the ones falling outside the confidence bands).

## References

Abramo, G., Cicero, T., D'Angelo, C.A. (2011). Assessing the varying level of impact measurement accuracy as a function of the citation window length. *Journal of Informetrics*, 5(4), 659-667.

Abramo, G., Cicero, T., D'Angelo, C.A. (2012a). Revisiting size effects in higher education research productivity. *Higher Education*, 63(6), 701-717.

Abramo, G., Cicero, T., D'Angelo, C.A. (2012c). The dispersion of research performance within and between universities as a potential indicator of the competitive intensity in higher education systems. *Journal of Informetrics*, 6(2), 155-168.

Abramo, G., D'Angelo, C.A. (2016a). A farewell to the MNCS and like size-independent indicators. *Journal of Informetrics*, 10(3), 646-651.

Abramo, G., D'Angelo, C.A. (2016b). A farewell to the MNCS and like size-independent indicators: Rejoinder. *Journal of Informetrics*, 10(3), 679-683.

Abramo, G., D'Angelo, C.A. (2014). How do you define and measure research productivity? *Scientometrics,* 101(2), 1129-1144.

Abramo, G., D'Angelo, C.A., Cicero, T. (2012b). What is the appropriate length of the publication period over which to assess research performance? *Scientometrics*, 93(3), 1005-1017.

Abramo, G., D'Angelo, C.A., Di Costa, F. (2011). A national-scale cross-time analysis of university research performance. *Scientometrics*, 87(2), 399-413.

Abramo, G., D'Angelo, C.A., Di Costa, F. (2014). Investigating returns to scope of research fields in universities. *Higher Education*, 68(1), 69-85.

Abramo, G., D'Angelo, C.A., Grilli, L. (2015). Funnel plots for visualizing uncertainty

in the research performance of institutions. *Journal of Informetrics*, 9(4), 954-961.

Abramo, G., D'Angelo, C.A., Rosati, F. (2013). The importance of accounting for the number of co-authors and their order when assessing research performance at the individual level in the life sciences. *Journal of Informetrics,* 7(1), 198–208.

Abramo, G., D'Angelo, C.A., Solazzi, M. (2010). National research assessment exercises: a measure of the distortion of performance rankings when labor input is treated as uniform. *Scientometrics*, 84(3), 605-619.

Box, G.E.P., Cox, D.R. (1964). An analysis of transformations. *Journal of the Royal Statistical Society, Series B*, 26, 211-252.

Claassen, C. (2015). Measuring university quality. *Scientometrics, 104*(3), 793-807.

D'Angelo, C.A., Giuffrida C., Abramo, G. (2011). A heuristic approach to author name disambiguation in bibliometrics databases for large-scale research assessments. *Journal of the American Society for Information Science and Technology*, 62(2), 257-269.

Hicks, D. (2012). Performance-based university research funding systems. *Research Policy*, 41(2), 251–261

Hicks, D., Wouters, P., Waltman, L., de Rijcke, S., Rafols, I. (2015). Bibliometrics: The Leiden Manifesto for research metrics. *Nature,* 520(7548), 429–431.

Ieva, F., Paganoni, A.M. (2015). Detecting and visualizing outliers in provider profiling via funnel plots and mixed effect models. *Health Care Management Science*, 18(2), 166-172.,.

Marginson, S., (1997). Steering from a distance: power relations in Australian higher education. *Higher Education*, 34(1), 63–80.

Moed, H.F. (2005). *Citation Analysis in Research Evaluation*. Springer, ISBN: 978-1-4020-3713-9

QS-Quacquarelli Symonds (2016). *World University Rankings 2015/6*. Retrieved from http://www.topuniversities.com/qs-world-university-rankings, July 7, 2016.

Royston, P. (1992). Approximating the shapiro-wilk W-test for non-normality. *Statistics and Computing*, 2(3), 117-119.

Scimago (2016). *Scimago institution ranking*. Retrieved from http://www.scimagoir.com/research.php, July 7, 2016.

SJTU-Shanghai Jiao Tong University (2016). *Academic Ranking of World Universities 2015*. Retrieved from: http://www.shanghairanking.com/ARWU2015.html, July 7, 2016.

Spiegelhalter, D.J. (2005). Funnel plots for comparing institutional performance. *Statistics in Medicine,* 24(8), 1185-1202.

THE-Times Higher Education (2016). *World University Rankings 2015-2016.* Retrieved from http://www.timeshighereducation.co.uk/world-university-rankings/2015-15/world-ranking, July 7, 2016.

Waltman, L., Calero-Medina, C., Kosten, J., Noyons, E.C.M., Tijssen, R.J.W., Van Eck, N.J., Van Leeuwen, T.N., Van Raan, A.F.J., Visser, M.S., Wouters, P. (2012). The Leiden Ranking 2011/2012: Data collection, indicators, and interpretation. *Journal of the American Society for Information Science and Technology*, 63(12), 2419–2432.

Waltman, L., Van Eck, N.J., Van Leeuwen, T.N., Visser, M.S., Van Raan, A.F.J. (2011). Towards a new crown indicator: Some theoretical considerations. *Journal of Informetrics*, 5(1), 37-47.